# Constructing Polar Codes Using Iterative Bit-Channel Upgrading

**Arash Ghayoori and T. Aaron Gulliver**

Department of Electrical and Computer Engineering,
University of Victoria,
Victoria, BC, Canada

agullive@ece.uvic.ca
arashgh@ece.uvic.ca

*Abstract*

The definition of polar codes given by Arikan is explicit, but the construction complexity is an issue. This is due to the exponential growth in the size of the output alphabet of the bit-channels as the codeword length increases. Tal and Vardy recently presented a method for constructing polar codes which controls this growth. They approximated each bit-channel with a "better" channel and a "worse" channel while reducing the alphabet size. They constructed a polar code based on the "worse" channel and used the "better" channel to measure the distance from the optimal channel. This paper considers the knowledge gained from the perspective of the "better" channel. A method is presented using iterative upgrading of the bit-channels which successively results in a channel closer to the original one. It is shown that this approach can be used to obtain a channel arbitrarily close to the original channel, and therefore to the optimal construction of a polar code.

*1. Introduction*

Polar codes [1] were recently introduced by Arikan for binary-input symmetric-output memoryless (BMS) channels. Since their introduction, they have attracted much attention because of the following favorable characteristics:

1. They have an explicit construction.
2. They are known to be capacity achieving.
3. They have efficient encoding and decoding algorithms.

Several generalizations of the work of Arikan have been made in terms of both the definition and application of polar codes, but in this paper, we only consider the original setting in [1].

Although the construction of polar codes presented in [1] is explicit, it is only proven to be efficient for the case of the binary erasure channel (BEC). There have been several attempts at an efficient construction for the general case, but with limited success. The first approach is that of Mori and Tanaka [2], [3], who considered using convolution to construct polar codes. The number of convolutions needed with their method is on the order of the code length. The difficulty with this approach lies in the fact that an exact implementation of the convolutions is very complex computationally and thus impractical. Tal and Vardy [4] introduced the idea of using quantization. They considered the fact that every channel is associated with a probability of error and used degrading and upgrading quantization to derive lower and upper bounds, respectively. Both quantizations result in a new channel with a smaller output alphabet. The degrading quantization results in a channel degraded with respect to the original one, while the upgrading quantization results in a channel upgraded with respect to the original one. They also showed that the approximations are very close to the original channel.

In this paper, we first review the concepts introduced in [4]. We then build upon them and extend the approximations made in [4]. In particular, iterative upgrading is introduced which successively provides a channel closer to the original channel. The resulting upgraded channel is very close to the original one, so that it can be used to directly construct polar codes.

*1.1 Polar Codes*

In this section, we briefly review polar codes. The notation follows that in [4]. A memoryless channel $W$ is used to transmit information. $X$ and $Y$ are the input and output alphabets, respectively, associated with the channel. This is denoted by $W: X \to Y$. $W(Y|X)$ is the probability of observing $y \in Y$ when $x \in X$ was transmitted through $W$. As mentioned in the introduction, we assume throughout that the input to $W$ is binary so that $X = \{0,1\}$. Further, $W$ is assumed to be a symmetric channel, so that there exists a permutation $\pi: Y \to Y$ such that:
i) $\pi^{-1} = \pi$, and
ii) $W(y|1) = W(\pi(y)|0)$.
For simplicity, $\pi(y)$ is abbreviated as $y'$. Finally, the output alphabet $Y$ of $W$ is assumed to be finite.

Let the codeword length be $n = 2^m$. For $y = (y_i)_{i=0}^{n-1}$ and $u = (u_i)_{i=0}^{n-1}$, let

$$W^n(y \mid u) = \prod_{i=0}^{n-1} W(y_i \mid u_i).$$

This means that $W^n$ corresponds to $n$ independent uses of the channel $W$. The polarization phenomenon introduced in [1] is that of transforming $n$ independent uses of the channel $W$ into $n$ bit-channels defined as follows. For $0 \leq i < n$, bit channel $W_i^{(m)}$ has a binary input alphabet $X$ and output alphabet $Y^n \times X^i$. Let $G$ be the kernel matrix [1]

$$G = \begin{pmatrix} 1 & 0 \\ 1 & 1 \end{pmatrix},$$

$G^{\otimes m}$ be the $m$-fold Kronecker product of $G$, and $B_m$ be the $n \times n$ bit-reversal matrix defined in [1]. Then, for input $u_i \in X$ and output $(y, u_0^{i-1}) \in Y^n \times X^i$, the probability $W_i^{(m)}((y, u_0^{i-1}) \mid u_i)$ is defined as

$$W_i^{(m)}\left((y, u_0^{i-1}) \mid u_i\right) = \frac{1}{2n-1} \sum_{v \in \{0,1\}^{n-1-i}} W^n\left(y \mid (u_0^{i-1}, u_i, v) B_m G^{\otimes m}\right).$$

Thus the problem of constructing a polar code of dimension $k$ becomes one of finding the $k$ "best" bit-channels. Two criteria can be followed for this purpose. In [1], it is suggested that the $k$ bit-channels with the lowest Bhattacharyya bounds be chosen, mainly because the bounds can easily be calculated. The second criterion [4] is more straightforward, as it ranks the channels according to the probability of misdecoding $u_i$ given the input $(y, u_0^{i-1})$. The latter criterion will be employed here.

Since the definition of a bit-channel in [1] is straightforward, it may appear that the construction of polar codes is simple. However, this is rarely the case. This is because the difficulty in constructing polar codes lies in the fact that the output alphabet size of each bit channel is exponential in the codeword length. Thus a direct use of the ranking criterion is only tractable for short codes. From [1], the increase in output alphabet size happens in stages. $W_{2i}^{(m+1)}$ and $W_{2i+1}^{(m+1)}$ can be constructed recursively from $W_i^{(m)}$ according to

$$W_{2i}^{(m+1)}(y_1, y_2 \mid u_1) = (W_i^{(m)} \otimes W_i^m)(y_1, y_2 \mid u_1) = \sum_{u_2 \in X} \frac{1}{2} W_i^{(m)}(y_1 \mid u_1 \oplus u_2) W_i^{(m)}(y_2 \mid u_2),$$

$$W_{2i+1}^{(m+1)}(y_1, y_2 \mid u_1) = (W_i^{(m)} \oplus W_i^m)(y_1, y_2, u_1 \mid u_2) = \frac{1}{2} W_i^{(m)}(y_1 \mid u_1 \oplus u_2) W_i^{(m)}(y_2 \mid u_2).$$

Thus going from stage $m$ to $m+1$ roughly squares the output alphabet size. As suggested in [4], this growth in the output alphabet $Y_i^{(m)}$ can be controlled by replacing the channel $W_i^{(m)}$ with an approximation. In [4], Tal and Vardy dealt with this issue by controlling the growth at each stage of polarization. They introduced two approximation methods, one transforming the channel into a "better" one and the other, transforming it into a "worse" one. We use their definitions for "degraded" and "upgraded" channels here.

***Definition 1:*** For channel $Q: X \to Z$ to be degraded with respect to $W: X \to Y$, an intermediate channel $P: Y \to Z$ must be found such that for all $z \in Z$ and $x \in X$

$$Q(z|x) = \sum_{y \in Y} W(y|x) P(z|y)$$

We write this as $Q \leq W$ ($Q$ is degraded with respect to $W$).

***Definition 2:*** For channel $Q': X \to Z'$ to be upgraded with respect to $W: X \to Y$, an intermediate channel $P: Z' \to Y$ must be found such that for all $z' \in Z'$ and $x \in X$

$$W(y|x) = \sum_{z' \in Z'} Q'(z'|x) P(y|z')$$

We write this as $Q' \geq W$ ($Q'$ is upgraded with respect to $W$).

The following results hold regarding the degrading and upgrading operations [4].

1. $W_1 \leq W_2$ if and only if $W_2 \geq W_1$.
2. If $W_1 \leq W_2$ and $W_2 \leq W_3$ then $W_1 \leq W_3$.
3. If $W_1 \geq W_2$ and $W_2 \geq W_3$ then $W_1 \geq W_3$.
4. $W \leq W$ and $W \geq W$.
5. If a channel $W_{eq}$ is both degraded and upgraded with respect to W, then W and $W_{eq}$ are equivalent. ($W_{eq} \equiv W$)
6. If $W_1 \equiv W_2$ and $W_2 \equiv W_3$ then $W_1 \equiv W_3$.
7. If $W_1 \equiv W_2$ then $W_2 \equiv W_1$.
8. $W \equiv W$

There are three important quantities with respect to the BMS channel $W: X \to Y$:

    i) The probability of error with Maximum Likelihood (ML) decoding, where ties are broken arbitrarily, and the input distribution is Bernoulli ($p(0) = p(1) = \frac{1}{2}$)

$$P_e(W) = \sum_{y \in Y: W(y|0) < W(y|1)} W(y|0) + \sum_{y \in Y: W(y|0) = W(y|1)} W(y|0)/2$$

    ii) The Bhattacharyya parameter

$$Z(W) = \sum_{y \in Y} \sqrt{W(y|0) W(y|1)}$$

iii) The channel capacity

$$I(W) = \sum_{y \in Y} \sum_{x \in X} \frac{1}{2} W(y|x) \log \frac{W(y|x)}{\frac{1}{2}W(y|0) + \frac{1}{2}W(y|1)}$$

The behavior of these quantities with respect to the degrading and upgrading relations is as follows:

9. *(Lemma 1 in [4])* Let $W: X \to Y$ be a BMS channel and let $Q: X \to Z$ be degraded with respect to $W$, that is, $Q \leq W$. Then

$$P_e(Q) \geq P_e(W)$$
$$Z(Q) \geq Z(W)$$
$$I(Q) \leq I(W)$$

These results also hold if "degraded" is replaced by "upgraded", and $\leq$ is replaced by $\geq$. Specifically, if $W \equiv Q$, then the inequalities are in fact equalities.

10. *(Lemma 3 in [4])* Fix a binary input channel $W: X \to Y$, and denote

$$W_\otimes = W \otimes W$$
$$W_\oplus = W \oplus W$$

Next, let $Q \leq W$ be a degraded channel with respect to $W$, and denote

$$Q_\otimes = Q \otimes Q$$
$$Q_\oplus = Q \oplus Q$$

Then, $Q_\otimes \leq W_\otimes$ and $Q_\oplus \leq W_\oplus$. Thus the degradation relation is preserved by the channel transformation operation. Moreover, these results hold if "degraded" is replaced by "upgraded" and $\leq$ is replaced by $\geq$.

*Assumptions:* As in [4], it is assumed that i) all channels are BMS and have no output symbols $y$ such that $y$ and $y'$ are equal, and ii) given a generic BMS channel $W: X \to Y$, for all $y \in Y$, at least one of the probabilities $W(y|0)$ and $W(y'|0)$ is positive. It was shown in [4] that these assumptions do not limit the results.

Based on the above assumptions, we now define an associated likelihood ratio for an output symbol [4].

***Definition 3:*** With each output symbol $y \in Y$ of the BMS channel $W: X \to Y$, a likelihood ratio, $LR_W(y)$, is associated which is defined as follows.

$$LR_W(y) = \frac{W(y|0)}{W(y'|0)}.$$

If $W(y'|0) = 0$, then define $LR_W(y) = \infty$.

If the channel $W$ is understood from the context, $LR_W(y)$ is abbreviated to $LR(y)$.

## 1.2 Algorithm Summary

In this section, a summary is given of the algorithms introduced in [4] for approximating a bit channel. Algorithms 1 and 2 present the procedures given by Tal and Vardy to obtain degraded and upgraded approximations of the bit channel $W_i^{(m)}$, respectively. That is, they provide a BMS channel that is degraded or upgraded with respect to $W_i^{(m)}$.

### *Algorithm 1: The degrading procedure introduced in [4]*

**Input:** A BMS channel $W$, a bound $\mu$ on the output alphabet size, and an index $i = <b_1, b_2, ..., b_m>_2$

**Output:** A BMS channel that is degraded with respect to the bit channel $W_i^{(m)}$

1. degrading_merge $(W, \mu) \to Q$
2. For $j = 1, 2, ..., m$ do
3.    If $b_j = 0$ then
4.       $Q \otimes Q \to W$
5.    else
6.       $Q \oplus Q \to W$
7.    degrading_merge $(W, \mu) \to Q$;
8. Return $Q$;

### *Algorithm 2: The upgrading procedure introduced in [4]*

**Input:** A BMS channel $W$, a bound $\mu$ on the output alphabet size, and an index $i = <b_1, b_2, ..., b_m>_2$

**Output:** A BMS channel that is upgraded with respect to the bit channel $W_i^{(m)}$

1. upgrading_merge $(W, \mu) \to Q$
2. For $j = 1, 2, ..., m$ do

3.     If $b_j = 0$ then
4.         $Q' \otimes Q' \rightarrow W$
5.     else
6.         $Q' \oplus Q' \rightarrow W$
7.     upgrading_merge $(W, \mu) \rightarrow Q$;
8. Return $Q$;

## 1.3 How to Construct Polar Codes

In general, when constructing a polar code, the following information is needed:
   i)     an underlying channel $W: X \rightarrow Y$,
   ii)     a specified codeword length $n = 2^m$ and
   iii)     a target block error rate $e_{Block}$.

The **ideal construction** of a polar code is as follows.

**Step 1.** Calculate the bit-misdecoding probabilities of all the bit-channels.

**Step 2.** Choose the largest possible subset of bit-channels such that the sum of their bit-misdecoding probabilities is less than or equal to the target block error rate.

**Step 3.** Span the resulting code by the rows in $B_m G^{\otimes m}$ corresponding to the bit-channels chosen in Step 2.

**Step 4.** The rate of this ideally designed polar code is denoted by $R_{exact}$.

The difficulty with the ideal design of a polar code lies in the first step above. Calculating the bit-misdecoding probabilities of bit-channels is a computationally complex task. Thus approximations of this step are desirable. With this in mind, a **practical construction** of a polar code can then be expressed as follows.

**Step 1.** Execute Algorithm 1 on the original channel to obtain a degraded version of it.

**Step 2.** Calculate the bit-misdecoding probabilities of the degraded bit-channels (since the output alphabet size of the degraded channels is limited, this is computationally tractable).

**Step 3.** Follow the steps in the ideal design of a polar code for the degraded bit-channels.

**Step 4.** Denote the rate of this code as $R_{degraded}$.

**Step 5.** Execute Algorithm 2 on the original channel to obtain an upgraded version of it.

**Step 6.** Calculate the bit-misdecoding probabilities of the upgraded bit-channels (since the output alphabet size of the upgraded channels is limited, this is computationally tractable).

**Step 7.** Follow the steps in the ideal design of a polar code for the upgraded bit-channels.

**Step 8.** Denote the rate of this code as $R_{upgraded}$.

**Step 9.** Consider the difference $R_{upgraded} - R_{degraded}$ to estimate the distance from the optimal design. Note that $R_{degraded} \leq R_{exact} \leq R_{upgraded}$.

## 1.4 Merging Functions

The alphabet size reducing degrading and upgrading functions referred to as degrading_merge and upgrading_merge in Algorithms 1 and 2 in [4] are defined as follows.

***Definition 4:*** For a BMS channel W and positive integer $\mu$, the output of degrading_merge $(W, \mu)$ is a BMS channel $Q$ such that
   i)   $Q$ is degraded with respect to $W$; and
   ii)  The size of the output alphabet of $Q$ is not greater than $\mu$.

***Definition 5:*** For a BMS channel W and positive integer $\mu$, the output of upgrading_merge $(W, \mu)$ is a BMS channel $Q$ such that
   i)   $Q$ is upgraded with respect to $W$; and
   ii)  The size of the output alphabet of $Q$ is not greater than $\mu$.

They are called merging functions because output symbols are merged in order to reduce the output alphabet size while producing degraded or upgraded version of the channel.

In [4], four lemmas were proven (Lemmas 5, 7, 9, 10) that are referred to in this paper as tools 1 to 4, respectively, for code construction. They are restated here as they will be used in the remainder of the paper.

***Tool 1 (Lemma 5 in [4]):*** Let $W : X \to Y$ be a BMS channel and let $y_1$ and $y_2$ be symbols in the output alphabet $Y$. Define the channel $Q : X \to Z$ as follows. The output alphabet $Z$ is given by

$$Z = Y \setminus \{y_1, y_1', y_2, y_2'\} \cup \{z_{12}, z_{12}'\}$$

For all $x \in X$ and $z \in Z$, define

$$Q(z \mid x) = \begin{cases} W(y_1 \mid x) + W(y_2 \mid x), & z = z_{12} \\ W(y_1' \mid x) + W(y_2' \mid x), & z = z_{12}' \\ W(z \mid x), & \text{otherwise} \end{cases}$$

Then $Q \le W$, that is, $Q$ is degraded with respect to $W$.

Tool 1 allows two consecutive output symbols to be merged together while reducing the output alphabet size, resulting in a degraded version of the original channel.

The degrading_merge function in [4] can be expressed as follows.
Consider a BMS channel with a specified output alphabet size. The goal is to reduce the channel alphabet size to the specified value while transforming the original channel into a degraded version of itself. The first step is to compare the output alphabet size with the desired output alphabet size. If the output alphabet size is not more than that desired, then take the given channel as its degraded version. Otherwise, a representative from each pair of output symbols is chosen with the same index such that its likelihood ratio is not less than one. The next step is to arrange these representatives in ascending order. Then the output alphabet index is found for which the mutual information of the channel resulting from applying Tool 1 to the original channel and the two consecutive symbols, starting with that index, is maximized. After finding this index, Tool 1 is applied to obtain a degraded channel with an alphabet size two symbols smaller than that of the original channel. The same process can be applied to the resulting degraded channel. This process is repeated until the desired output alphabet size is obtained.

***Tool 2 (Lemma 7 in [4]):*** Let $W: X \to Y$ be a BMS channel and let $y_1$ and $y_2$ be symbols in the output alphabet $Y$. Denote $\lambda_2 = LR(y_2)$ and $\lambda_1 = LR(y_1)$. Assume that

$$1 \leq \lambda_1 \leq \lambda_2.$$

Further, let $a_1 = W(y_1 | 0)$ and $b_1 = W(y_1' | 0)$. Define $\alpha_2$ and $\beta_2$ as follows

$$\alpha_2 = \frac{\lambda_2(a_1 + b_1)}{\lambda_2 + 1}$$

$$\beta_2 = \frac{a_1 + b_1}{\lambda_2 + 1}$$

and for real numbers $\alpha, \beta$ and $x \in X$, define

$$t(\alpha, \beta | x) = \begin{Bmatrix} \alpha, x = 0 \\ \beta, x = 1 \end{Bmatrix}$$

Define the channel $Q': X \to Z'$ as follows.

The output alphabet $Z'$ is given by

$$Z' = Y \setminus \{y_2, y_2', y_1, y_1'\} \cup \{z_2, z_2'\}$$

For all $x \in X$ and $z \in Z'$

$$Q'(z|x) = \begin{Bmatrix} W(y_2 | x) + t(\alpha_2, \beta_2 | x), z = z_2 \\ W(y_2' | x) + t(\alpha_2, \beta_2 | x), z = z_2' \\ W(z|x), \text{ otherwise} \end{Bmatrix}$$

Then $Q' \geq W$, that is, $Q'$ is upgraded with respect to $W$.

Tool 2 allows us to take two consecutive output symbols, merge them together to reduce the output alphabet size and obtain an upgraded version of the original channel.

**Tool 3 (Lemma 9 in [4]):** Let $W: X \to Y$ be a BMS channel, and let $y_1, y_2$ and $y_3$ be symbols in the output alphabet $Y$. Denote $\lambda_1 = LR(y_1), \lambda_2 = LR(y_2)$ and $\lambda_3 = LR(y_3)$. Assume that

$$1 \le \lambda_1 < \lambda_2 < \lambda_3.$$

Next, let $a_2 = W(y_2 | 0)$ and $b_2 = W(y_2' | 0)$. Define $\alpha_1, \beta_1, \alpha_3$ and $\beta_3$ as follows.

$$\alpha_1 = \frac{\lambda_1(\lambda_3 b_2 - a_2)}{\lambda_3 - \lambda_1}$$

$$\beta_1 = \frac{\lambda_3 b_2 - a_2}{\lambda_3 - \lambda_1}$$

$$\alpha_3 = \frac{\lambda_3(a_2 - \lambda_1 b_2)}{\lambda_3 - \lambda_1}$$

$$\beta_3 = \frac{a_2 - \lambda_1 b_2}{\lambda_3 - \lambda_1}$$

and for real numbers $\alpha, \beta$ and $x \in X$, define

$$t(\alpha, \beta | x) = \begin{cases} \alpha, & x = 0 \\ \beta, & x = 1 \end{cases}$$

Define the channel $Q': X \to Z'$ as follows.
The output alphabet $Z'$ is given by

$$Z' = Y \setminus \{y_1, y_1', y_2, y_2', y_3, y_3'\} \cup \{z_1, z_1', z_2, z_2'\}$$

For all $x \in X$ and $z \in Z'$, define

$$Q'(z | x) = \begin{cases} W(y_1 | x) + t(\alpha_1, \beta_1 | x), & z = z_1 \\ W(y_1' | x) + t(\alpha_1, \beta_1 | x), & z = z_1' \\ W(y_3 | x) + t(\alpha_3, \beta_3 | x), & z = z_3 \\ W(y_3' | x) + t(\alpha_3, \beta_3 | x), & z = z_3' \\ W(z | x), & \text{otherwise} \end{cases}$$

Then $Q' \ge W$, that is, $Q'$ is upgraded with respect to $W$.

With Tool 3, three consecutive output symbols can be merged together to reduce the output alphabet size and obtain an upgraded version of the original channel.

The upgrading_merge function in [4] can be described as follows.
Assume a BMS channel with a specified output alphabet size. It is desired to reduce the alphabet size to this value while transforming the original channel into an upgraded version of itself. First, the output alphabet size is compared with the desired output alphabet size. If the output alphabet size is not more than the desired size, the channel itself is taken as the upgraded version. Otherwise, as in the merge-degrading procedure, choose a representative from each pair of output symbols with the same index such that its likelihood ratio is not less than one. Arrange them according to their likelihood ratio values in ascending order. Next, for a parameter $\varepsilon$, check if there exists an output alphabet index such that the division of the likelihood ratios of two consecutive symbols is less than $1+\varepsilon$. If so, apply Tool 2 repeatedly until no such index exists. Now, the main step is to find the index for which the mutual information of the channel resulting from applying Tool 3 to the original channel and three consecutive symbols, starting with that index, is maximized. After finding this index, Tool 3 is applied to obtain an upgraded channel with an alphabet size that is two symbols less than that of the original channel. This process is applied repeatedly on the resulting upgraded channels until the desired output alphabet size is obtained.

**Tool 4 (Lemma 10 in [4]):** Let $W$, $y_1$, $y_2$ and $y_3$ be as in Tool 3. Denote by $Q'_{123}: X \to Z'_{123}$ the result of applying Tool 3 to $W, y_1, y_2$ and $y_3$. Next, denote by $Q'_{23}: X \to Z'_{23}$ the result of applying Tool 2 to $W, y_2$ and $y_3$. Then $Q'_{23} \geq Q'_{123} \geq W$. $Q'_{123}$ can be considered a more faithful representation of $W$ than $Q'_{23}$ is.

Tool 4 hints that by increasing the number of symbols merged together to get an upgraded version of the channel, while reducing the output alphabet size, better channels can be obtained (the upgraded channel rate approaches the exact rate). This motivates the development of an algorithm for merging 4 symbols (Lemma 1), 5 symbols (Lemma 3) and 6 symbols (Lemma 5), to obtain upgraded versions of the channel. The key result is that merging more symbols provides a channel closer to the original channel. This leads to a general algorithm for merging symbols.

**Tool 5 (Theorems 6 and 11 in [4]):** No generality is lost by only considering merging of consecutive symbols and essentially nothing is gained by considering non-consecutive symbols.

Tool 5 indicates that only consecutive symbols need to be considered. This reduces the complexity of the proofs.

## *2. Relationships to Previous Results in Polar Codes*

In [4], Tal and Vardy established a framework for the efficient construction of polar codes. In their algorithm, when the number of output symbols of the bit-channel is more than the desired number, the bit-channel is approximated with both a worse channel and a better channel. They then constructed a polar code based on the worse channel while measuring the distance from the optimal result by considering the better channel.

The original channel lies between the better channel $Q'$ and the worse channel $Q$

$$Q \leq W \leq Q'$$

The focus in [4] was primarily on the worse channel $Q$. In this paper, we shift this paradigm and consider the original channel from the perspective of the upper bound (better channel) $Q'$. By increasing the number of symbols being merged together, more knowledge is gained regarding the original channel $W$ and the optimal construction of a polar code associated with this channel. We show that it is possible to get arbitrarily close to the original channel and therefore to the construction of an optimal polar code, which is the main contribution of this paper.

The remainder of this paper is organized as follows. In Lemma 1, we introduce a method to merge 4 symbols to obtain an upgraded version of the channel, while in Lemma 2 we prove that this results in a better channel than that using Lemma 9 in [4]. In Lemmas 3 and 5, we expand our method to 5 and 6 symbols, respectively, and in Lemmas 4 and 6 we prove that better channels are obtained using Lemmas 3 and 5, respectively. Finally, our approach is extended to a general algorithm for an arbitrary number of symbols.

**Lemma 1:** Let $W: X \to Y$ be a BMS channel and let $y_1, y_2, y_3$ and $y_4$ be symbols in the output alphabet $Y$. Denote $\lambda_1 = LR(y_1), \lambda_2 = LR(y_2), \lambda_3 = LR(y_3)$ and $\lambda_4 = LR(y_4)$. Assume that

$$1 \leq \lambda_1 < \lambda_2 \leq \lambda_3 < \lambda_4.$$

Next, let $a_2 = W(y_2 | 0), b_2 = W(y_2' | 0), a_3 = W(y_3 | 0)$ and $b_3 = W(y_3' | 0)$. Define $a, b, \alpha_1, \beta_1, \alpha_4$ and $\beta_4$ as follows.

$$a = a_2 + a_3$$
$$b = b_2 + b_3$$
$$\alpha_1 = \frac{\lambda_1(\lambda_4 b - a)}{\lambda_4 - \lambda_1}$$
$$\beta_1 = \frac{\lambda_4 b - a}{\lambda_4 - \lambda_1}$$

$$\alpha_4 = \frac{\lambda_4(a - \lambda_1 b)}{\lambda_4 - \lambda_1}$$

$$\beta_4 = \frac{a - \lambda_1 b}{\lambda_4 - \lambda_1}$$

and for real numbers $\alpha, \beta$ and $x \in X$, define

$$t(\alpha, \beta \mid x) = \begin{cases} \alpha, x = 0 \\ \beta, x = 1 \end{cases}$$

The BMS channel $Q': X \to Z'$ is defined as follows.

The output alphabet $Z'$ is given by

$$Z' = Y \setminus \{y_1, y_1', y_2, y_2', y_3, y_3', y_4, y_4'\} \cup \{z_1, z_1', z_4, z_4'\}$$

For all $x \in X$ and $z \in Z'$, define

$$Q'(z \mid x) = \begin{cases} W(y_1 \mid x) + t(\alpha_1, \beta_1 \mid x), z = z_1 \\ W(y_1' \mid x) + t(\alpha_1, \beta_1 \mid x), z = z_1' \\ W(y_4 \mid x) + t(\alpha_4, \beta_4 \mid x), z = z_4 \\ W(y_4' \mid x) + t(\alpha_4, \beta_4 \mid x), z = z_4' \\ W(z \mid x), \quad \text{otherwise} \end{cases}$$

Then $Q' \geq W$, that is, $Q'$ is upgraded with respect to $W$.

∎

***Proof:*** From Definition 2, for $Q'$ to be upgraded with respect to $W$, an intermediate channel $P: Z' \to Y$ must be found such that:
$$W(Y \mid X) = Q'(Z' \mid X) P(Y \mid Z') \qquad \text{(I)}$$

The transition probability matrix for $W$ is

$$W(Y \mid X) = \begin{pmatrix} a_1 & a_2 & a_3 & a_4 & b_4 & b_3 & b_2 & b_1 \\ b_1 & b_2 & b_3 & b_4 & a_4 & a_3 & a_2 & a_1 \end{pmatrix}$$

Considering the definition of $Q'$ in the lemma, the transition probability matrix for $Q'$ can be calculated as

$$Q'(Z'|X) = \begin{pmatrix} a_1+\alpha_1 & a_4+\alpha_4 & b_4+\alpha_4 & b_1+\alpha_4 \\ b_1+\beta_1 & a_4+\beta_4 & b_4+\beta_4 & a_1+\beta_1 \end{pmatrix}$$

$Q'$ is a 2×4 matrix and $W$ is a 2×8 matrix. Therefore, according to (I), the transition probability matrix for the intermediate channel $P$ must be a 4×8 matrix, and in general this matrix can be assumed to be

$$P(Y|Z') = \begin{pmatrix} p_1 & q_4 & q_2 & 0 & 0 & 0 & 0 & 0 \\ 0 & q_3 & q_1 & p_4 & 0 & 0 & 0 & 0 \\ 0 & 0 & 0 & 0 & p_4 & q_1 & q_3 & 0 \\ 0 & 0 & 0 & 0 & 0 & q_2 & q_4 & p_1 \end{pmatrix}$$

Substituting this into (I) gives the following equations that can be used to calculate $p_1, p_4, q_1, q_2, q_3$ and $q_4$

$$p_1(a_1+\alpha_1) = a_1$$
$$p_1(b_1+\alpha_4) = b_1$$
$$p_1(b_1+\beta_1) = b_1$$
$$p_1(a_1+\beta_1) = a_1$$
$$p_4(a_4+\alpha_4) = a_4$$
$$p_4(b_4+\alpha_4) = b_4$$
$$p_4(a_4+\beta_4) = b_4$$
$$p_4(b_4+\beta_4) = a_4$$

$$(a_1+\alpha_1)q_4 + (a_4+\alpha_4)q_3 = a_2$$
$$(b_1+\beta_1)q_4 + (a_4+\beta_4)q_3 = b_2$$
$$(b_4+\alpha_4)q_3 + (b_1+\alpha_4)q_4 = b_2$$
$$(b_4+\beta_4)q_3 + (a_1+\beta_1)q_4 = a_2$$
$$(a_1+\alpha_1)q_2 + (a_4+\alpha_4)q_1 = a_3$$
$$(b_1+\beta_1)q_2 + (a_4+\beta_4)q_1 = b_3$$
$$(b_4+\alpha_4)q_1 + (b_1+\alpha_4)q_2 = b_3$$
$$(b_4+\beta_4)q_1 + (a_1+\beta_1)q_2 = a_3$$

Solving these equations gives

$$p_1 = a_1 / (a_1 + \alpha_1) = b_1 / (b_1 + \alpha_4) = b_1 / (b_1 + \beta_1) = a_1 / (a_1 + \beta_1)$$

$$p_4 = a_4 / (a_4 + \alpha_4) = b_4 / (b_4 + \alpha_4) = b_4 / (a_4 + \beta_4) = a_4 / (b_4 + \beta_4)$$

$$q_2 = q_3 = 0$$

$$q_1 = a_3 / (a_4 + \alpha_4) = b_3 / (a_4 + \beta_4) = b_3 / (b_4 + \alpha_4) = a_3 / (b_4 + \beta_4)$$

$$q_4 = a_2 / (a_1 + \alpha_1) = b_2 / (b_1 + \beta_1) = b_2 / (b_1 + \alpha_4) = a_2 / (a_1 + \beta_1)$$

The resulting transition probability matrix $P$ satisfies the conditions to be a channel, which completes the proof that $Q'$ is upgraded with respect to $W$.

**Lemma 2:** Let $W, y_1, y_2, y_3$ and $y_4$ be as in Lemma 1. Denote by $Q'_{1234} : X \to Z'_{1234}$ the result of applying Lemma 1 to $W, y_1, y_2, y_3$ and $y_4$. Next, denote by $Q'_{234} : X \to Z'_{234}$ the result of applying Lemma 9 in [4] to $W, y_2, y_3$ and $y_4$. Then $Q'_{234} \succeq Q'_{1234} \succeq W$.

∎

**Proof:** From Definition 1, for $Q'_{1234}$ to be degraded with respect to $Q'_{234}$, an intermediate channel $P : Z'_{234} \to Z'_{1234}$ must be found such that

$$Q'_{1234}(Z_{1234} | X) = Q'_{234}(Z_{234} | X) P(Z_{1234} | Z_{234}) \qquad (II)$$

Recall that the two alphabets $Z'_{1234}$ and $Z'_{234}$ satisfy

$$Z'_{1234} = \{z_1, z_4, z_1', z_4'\} \cup A$$

$$Z'_{234} = \{z_2, z_4, z_2', z_4', y_1, y_1'\} \cup A$$

where $A = Y \setminus \{y_1, y_1', y_2, y_2', y_3, y_3', y_4, y_4'\}$ is the set of symbols not participating in either merge operation.

As determined previously, the transition probability matrix for $Q'_{1234}$ is

$$Q'_{1234}(Z_{1234} | X) = \begin{pmatrix} a_1 + \alpha_1 & a_4 + \alpha_4 & b_4 + \alpha_4 & b_1 + \alpha_4 \\ b_1 + \beta_1 & a_4 + \beta_4 & b_4 + \beta_4 & a_1 + \beta_1 \end{pmatrix}$$

Using the definition of $Q'_{234}$ introduced in Lemma 9 in [4], the transition probability matrix for channel $Q'_{234}$ is

$$Q'_{234}(Z_{234} | X) = \begin{pmatrix} a_2 + \alpha'_2 & a_4 + \alpha'_4 & a_1 & b_1 & b_4 + \alpha'_4 & b_2 + \alpha'_2 \\ b_2 + \beta'_2 & b_4 + \beta'_4 & b_1 & a_1 & a_4 + \beta'_4 & a_2 + \beta'_2 \end{pmatrix}$$

$Q'_{1234}$ is a 2×4 matrix and $Q'_{234}$ is a 2×6 matrix. Thus, according to (II) the transition probability matrix for the intermediate channel $P$ must be a 6×4 matrix, and in general this matrix can be assumed to have the form

$$P(Z_{1234} | Z_{234}) = \begin{pmatrix} p & 0 & 0 & 0 \\ 0 & 1 & 0 & 0 \\ q & 0 & 0 & 0 \\ 0 & 0 & 0 & q \\ 0 & 0 & 1 & 0 \\ 0 & 0 & 0 & p \end{pmatrix}$$

Substituting this into (II) gives the following equations which can be used to determine $p$ and $q$

$$a_4 + \alpha_4 = a_4 + \alpha'_4$$
$$b_4 + \alpha_4 = b_4 + \alpha'_4$$
$$b_4 + \beta'_4 = \beta_4 + a_4$$
$$b_4 + \beta_4 = a_4 + \beta'_4$$
$$(a_2 + \alpha'_2)p + qa_1 = a_1 + \alpha_1$$
$$(b_2 + \beta'_2)p + qb_1 = b_1 + \beta_1$$
$$(b_2 + \alpha'_2)p + qb_1 = b_1 + \alpha_4$$
$$(a_2 + \beta'_2)p + qa_1 = a_1 + \beta_1$$

Note that $\alpha_1, \beta_1, \alpha_4$ and $\beta_4$ are as defined in Lemma 1, but $\alpha'_2, \beta'_2, \alpha'_4$ and $\beta'_4$ are as defined in Lemma 9 in [4], which are

$$\alpha'_2 = \frac{\lambda_2(\lambda_4 b_3 - a_3)}{\lambda_4 - \lambda_2}$$

$$\beta'_2 = \frac{\lambda_4 b_3 - a_3}{\lambda_4 - \lambda_2}$$

$$\alpha'_4 = \frac{\lambda_4(a_3 - \lambda_2 b_3)}{\lambda_4 - \lambda_2}$$

$$\beta'_4 = \frac{a_3 - \lambda_2 b_3}{\lambda_4 - \lambda_2}$$

Solving these equations gives

$q = 0$

$$p = \frac{a_1 + \alpha_1}{a_2 + \alpha'_2} = \frac{b_1 + \beta_1}{b_2 + \beta'_2} = \frac{b_1 + \alpha_4}{b_2 + \alpha'_2} = \frac{a_1 + \beta_1}{a_2 + \beta'_2}$$

Choosing $p = 1$ results in a transition probability matrix that satisfies the requirements to be a channel, which completes the proof that $Q'_{1234}$ is degraded with respect to $Q'_{234}$ and so is a better approximation for channel $W$.

***Lemma 3:*** Let $W: X \to Y$ be a BMS channel and let $y_1, y_2, y_3, y_4$ and $y_5$ be symbols in the output alphabet $Y$. Denote $\lambda_1 = LR(y_1), \lambda_2 = LR(y_2), \lambda_3 = LR(y_3), \lambda_4 = LR(y_4)$ and $\lambda_5 = LR(y_5)$. Assume that

$$1 \le \lambda_1 < \lambda_2 \le \lambda_3 \le \lambda_4 < \lambda_5.$$

Next, let $a_2 = W(y_2 | 0), b_2 = W(y_2' | 0), a_3 = W(y_3 | 0), b_3 = W(y_3' | 0), a_4 = W(y_4 | 0)$ and $b_4 = W(y_4' | 0)$. Define $a, b, \alpha_1, \beta_1, \alpha_3, \beta_3, \alpha_5$ and $\beta_5$ as follows.

$$\alpha_3 = \frac{\lambda_3(a_2 + b_2)}{\lambda_3 + 1}$$

$$\beta_3 = \frac{a_2 + b_2}{\lambda_3 + 1}$$

$$a = a_3 + a_4 + \alpha_3$$

$$b = b_3 + \beta_3 + b_4$$

$$\alpha_1 = \frac{\lambda_1(\lambda_5 b - a)}{\lambda_5 - \lambda_1}$$

$$\beta_1 = \frac{\lambda_5 b - a}{\lambda_5 - \lambda_1}$$

$$\alpha_5 = \frac{\lambda_5(a - \lambda_1 b)}{\lambda_5 - \lambda_1}$$

$$\beta_5 = \frac{a - \lambda_1 b}{\lambda_5 - \lambda_1}$$

and for real numbers $\alpha, \beta$ and $x \in X$, define

$$t(\alpha, \beta \mid x) = \begin{cases} \alpha, x = 0 \\ \beta, x = 1 \end{cases}$$

Define the BMS channel $Q': X \to Z'$ as follows.

The output alphabet $Z'$ is given by

$$Z' = Y \setminus \{y_1, y_1', y_2, y_2', y_3, y_3', y_4, y_4', y_5, y_5'\} \cup \{z_1, z_1', z_5, z_5'\}$$

For all $x \in X$ and $z \in Z'$, define

$$Q'(z \mid x) = \begin{cases} W(y_1 \mid x) + t(\alpha_1, \beta_1 \mid x), z = z_1 \\ W(y_1' \mid x) + t(\alpha_1, \beta_1 \mid x), z = z_1' \\ W(y_5 \mid x) + t(\alpha_5, \beta_5 \mid x), z = z_5 \\ W(y_5' \mid x) + t(\alpha_5, \beta_5 \mid x), z = z_5' \\ W(z \mid x), \quad \text{otherwise} \end{cases}$$

Then $Q' \geq W$, that is, $Q'$ is upgraded with respect to $W$.

∎

***Proof:*** Based on Definition 2, for $Q'$ to be upgraded with respect to $W$, an intermediate channel $P: Z' \to Y$ must be found such that

$$W(Y \mid X) = Q'(Z' \mid X) P(Y \mid Z') \qquad \text{(III)}$$

The transition probability matrix for $W$ is

$$W(Y \mid X) = \begin{pmatrix} a_1 & a_2 & a_3 & a_4 & a_5 & b_5 & b_4 & b_3 & b_2 & b_1 \\ b_1 & b_2 & b_3 & b_4 & b_5 & a_5 & a_4 & a_3 & a_2 & a_1 \end{pmatrix}.$$

Considering the definition of $Q'$ in the lemma, the transition probability matrix for $Q'$ can be expressed as

$$Q'(Z' \mid X) = \begin{pmatrix} a_1 + \alpha_1 & a_5 + \alpha_5 & b_5 + \alpha_5 & b_1 + \alpha_5 \\ b_1 + \beta_1 & a_5 + \beta_5 & b_5 + \beta_5 & a_1 + \beta_1 \end{pmatrix}.$$

$Q'$ is a 2×4 matrix and $W$ is a 2×10 matrix. Thus, according to (III), the transition probability matrix for the intermediate channel $P$ must be a 4×10 matrix, and in general this matrix can be expressed as

$$P(Y|Z') = \begin{pmatrix} p_1 & p_2 & p_3 & p_4 & p_5 & 0 & 0 & 0 & 0 & 0 \\ q_1 & q_2 & q_3 & q_4 & q_5 & 0 & 0 & 0 & 0 & 0 \\ 0 & 0 & 0 & 0 & 0 & q_5 & q_4 & q_3 & q_2 & q_1 \\ 0 & 0 & 0 & 0 & 0 & p_5 & p_4 & p_3 & p_2 & p_1 \end{pmatrix}$$

Substituting this result into (III) gives the following equations from which $p_1, p_2, p_3, p_4, p_5, q_1, q_2, q_3, q_4$ and $q_5$ can be calculated

$$a_1 = (a_1 + \alpha_1)p_1 + (a_5 + \alpha_5)q_1$$
$$b_1 = (b_1 + \beta_1)p_1 + (a_5 + \beta_5)q_1$$
$$b_1 = (b_5 + \alpha_5)q_1 + (b_1 + \alpha_5)p_1$$
$$a_1 = (b_5 + \beta_5)q_1 + (a_1 + \beta_1)p_1$$
$$a_2 = (a_1 + \alpha_1)p_2 + (a_5 + \alpha_5)q_2$$
$$b_2 = (b_1 + \beta_1)p_2 + (a_5 + \beta_5)q_2$$
$$b_2 = (b_5 + \alpha_5)q_2 + (b_1 + \alpha_5)p_2$$
$$a_2 = (b_5 + \beta_5)q_2 + (a_1 + \beta_1)p_2$$
$$a_3 = (a_1 + \alpha_1)p_3 + (a_5 + \alpha_5)q_3$$
$$b_3 = (b_1 + \beta_1)p_3 + (a_5 + \beta_5)q_3$$
$$b_3 = (b_5 + \alpha_5)q_3 + (b_1 + \alpha_5)p_3$$
$$a_3 = (b_5 + \beta_5)q_3 + (a_1 + \beta_1)p_3$$
$$a_4 = (a_1 + \alpha_1)p_4 + (a_5 + \alpha_5)q_4$$
$$b_4 = (b_1 + \beta_1)p_4 + (a_5 + \beta_5)q_4$$
$$b_4 = (b_5 + \alpha_5)q_4 + (b_1 + \alpha_5)p_4$$
$$a_4 = (b_5 + \beta_5)q_4 + (a_1 + \beta_1)p_4$$
$$a_5 = (a_1 + \alpha_1)p_5 + (a_5 + \alpha_5)q_5$$
$$b_5 = (b_1 + \beta_1)p_5 + (a_5 + \beta_5)q_5$$
$$b_5 = (b_5 + \alpha_5)q_5 + (b_1 + \alpha_5)p_5$$
$$a_5 = (b_5 + \beta_5)q_5 + (a_1 + \beta_1)p_5$$

Solving these equations gives

$$p_1 = a_1 / (a_1 + \alpha_1) = b_1 / (b_1 + \beta_1) = b_1 / (b_1 + \alpha_5) = a_1 / (a_1 + \beta_1)$$

$$q_1 = q_2 = 0$$

$$p_2 = a_2 / (a_1 + \alpha_1) = b_2 / (b_1 + \beta_1) = b_2 / (b_1 + \alpha_5) = a_2 / (a_1 + \beta_1)$$

$$p_3 = p_4 = 0$$

$$q_3 = a_3 / (a_5 + \alpha_5) = b_3 / (a_5 + \beta_5) = b_3 / (b_5 + \alpha_5) = a_3 / (b_5 + \beta_5)$$

$$q_4 = a_4 / (a_5 + \alpha_5) = b_4 / (a_5 + \beta_5) = b_4 / (b_5 + \alpha_5) = a_4 / (b_5 + \beta_5)$$

$$q_5 = 0$$

$$p_5 = a_5 / (a_1 + \alpha_1) = b_5 / (b_1 + \beta_1) = b_5 / (b_1 + \alpha_5) = a_5 / (a_1 + \beta_1)$$

The resulting transition probability matrix $P$ satisfies the conditions to be a channel, which completes the proof that $Q'$ is upgraded with respect to $W$.

**Lemma 4:** Let $W, y_1, y_2, y_3, y_4$ and $y_5$ be as in Lemma 3. Denote by $Q'_{12345} : X \rightarrow Z'_{12345}$ the result of applying Lemma 3 to $W, y_1, y_2, y_3, y_4$ and $y_5$. Next, denote by $Q'_{2345} : X \rightarrow Z'_{2345}$ the result of applying Lemma 1 to $W, y_2, y_3, y_4$ and $y_5$. Then $Q'_{2345} \geq Q'_{12345} \geq W$.

∎

**Proof:** From Definition 1, for $Q'_{12345}$ to be degraded with respect to $Q'_{2345}$, an intermediate channel $P : Z'_{2345} \rightarrow Z'_{12345}$ must be found such that

$$Q'_{12345}(Z_{12345} | X) = Q'_{2345}(Z_{2345} | X) P(Z_{12345} | Z_{2345}) \qquad \text{(IV)}$$

Recall that the two alphabets $Z'_{12345}$ and $Z'_{2345}$ satisfy

$$Z'_{12345} = \{z_1, z_5, z_1', z_5'\} \cup A$$

$$Z'_{2345} = \{z_2, z_5, z_2', z_5', y_1, y_1'\} \cup A$$

where $A = Y \setminus \{y_1, y_1', y_2, y_2', y_3, y_3', y_4, y_4', y_5, y_5'\}$ is the set of symbols not participating in either merge operation.

As calculated previously, the transition probability matrix for the channel $Q'_{12345}$ is

$$Q'_{12345}(Z_{12345} | X) = \begin{pmatrix} a_1 + \alpha_1 & a_5 + \alpha_5 & b_5 + \alpha_5 & b_1 + \alpha_5 \\ b_1 + \beta_1 & a_5 + \beta_5 & b_5 + \beta_5 & a_1 + \beta_1 \end{pmatrix}$$

From the definition of $Q'_{2345}$ introduced in Lemma 1, the transition probability matrix for channel $Q'_{2345}$ is

$$Q'_{2345}(Z_{2345}|X) = \begin{pmatrix} a_2+\alpha'_2 & a_5+\alpha'_5 & a_1 & b_1 & b_5+\alpha'_5 & b_2+\alpha'_2 \\ b_2+\beta'_2 & b_5+\beta'_5 & b_1 & a_1 & a_5+\beta'_5 & a_2+\beta'_2 \end{pmatrix}$$

$Q'_{12345}$ is a 2×4 matrix and $Q'_{2345}$ is a 2×6 matrix. Thus, according to (IV), the transition probability matrix for the intermediate channel $P$ must be a 6×4 matrix, and in general this matrix can be assumed to be

$$P(Z_{12345}|Z_{2345}) = \begin{pmatrix} p & 0 & 0 & 0 \\ 0 & 1 & 0 & 0 \\ q & 0 & 0 & 0 \\ 0 & 0 & 0 & q \\ 0 & 0 & 1 & 0 \\ 0 & 0 & 0 & p \end{pmatrix}$$

Substituting this into (IV), gives the following equations from which $p$ and $q$ can be determined

$$a_5 + \alpha_5 = a_5 + \alpha'_5$$
$$b_5 + \alpha_5 = b_5 + \alpha'_5$$
$$b_5 + \beta'_5 = \beta_5 + a_5$$
$$b_5 + \beta_5 = a_5 + \beta'_5$$
$$(a_2 + \alpha'_2)p + qa_1 = a_1 + \alpha_1$$
$$(b_2 + \beta'_2)p + qb_1 = b_1 + \beta_1$$
$$(b_2 + \alpha'_2)p + qb_1 = b_1 + \alpha_5$$

Note that $\alpha'_2, \beta'_2, \alpha'_5$ and $\beta'_5$ are as defined in Lemma 1, which is

$$\alpha'_2 = \frac{\lambda_2(\lambda_5 b - a)}{\lambda_5 - \lambda_2}$$

$$\beta'_2 = \frac{\lambda_5 b - a}{\lambda_5 - \lambda_2}$$

$$\alpha'_5 = \frac{\lambda_5(a - \lambda_2 b)}{\lambda_5 - \lambda_2}$$

$$a = a_3 + a_4$$
$$b = b_3 + b_4$$

Solving these equations gives

$$q = 0$$

$$p = \frac{a_1 + \alpha_1}{a_2 + \alpha'_2} = \frac{b_1 + \beta_1}{b_2 + \beta'_2} = \frac{b_1 + \alpha_5}{b_2 + \alpha'_2} = \frac{a_1 + \beta_1}{a_2 + \beta'_2}$$

Choosing $p = 1$ results in a transition probability matrix that satisfies the conditions for a channel, which completes the proof that $Q'_{12345}$ is degraded with respect to $Q'_{2345}$ and so is a better approximation for channel $W$.

**Lemma 5:** Let $W : X \to Y$ be a BMS channel and let $y_1, y_2, y_3, y_4, y_5$ and $y_6$ be symbols in the output alphabet $Y$. Denote $\lambda_1 = LR(y_1), \lambda_2 = LR(y_2), \lambda_3 = LR(y_3), \lambda_4 = LR(y_4), \lambda_5 = LR(y_5)$ and $\lambda_6 = LR(y_6)$. Assume that

$$1 \leq \lambda_1 < \lambda_2 \leq \lambda_3 \leq \lambda_4 \leq \lambda_5 < \lambda_6.$$

Next, let $a_2 = W(y_2|0), b_2 = W(y_2'|0), a_3 = W(y_3|0), b_3 = W(y_3'|0), a_4 = W(y_4|0)$, $a_5 = W(y_5|0)$ and $b_5 = W(y_5'|0)$. Define $a, b, a', b', \alpha_1, \beta_1, \alpha_4, \beta_4, \alpha_6$ and $\beta_6$ as follows

$$a = a_2 + a_3$$

$$b = b_2 + b_3$$

$$\alpha_4 = \frac{\lambda_4(a+b)}{\lambda_4 + 1}$$

$$\beta_4 = \frac{a+b}{\lambda_4 + 1}$$

$$a' = a_4 + a_5 + \alpha_4$$

$$b' = b_4 + b_5 + \beta_4$$

$$\alpha_1 = \frac{\lambda_1(\lambda_6 b' - a')}{\lambda_6 - \lambda_1}$$

$$\beta_1 = \frac{\lambda_6 b' - a'}{\lambda_6 - \lambda_1}$$

$$\alpha_6 = \frac{\lambda_6(a' - \lambda_1 b')}{\lambda_6 - \lambda_1}$$

$$\beta_6 = \frac{a' - \lambda_1 b'}{\lambda_6 - \lambda_1}$$

and for real numbers $\alpha, \beta$ and $x \in X$, define

$$t(\alpha, \beta \mid x) = \begin{cases} \alpha, x = 0 \\ \beta, x = 1 \end{cases}$$

Define the BMS channel $Q': X \to Z'$ as follows.

The output alphabet $Z'$ is given by

$$Z' = Y \setminus \{y_1, y_1', y_2, y_2', y_3, y_3', y_4, y_4', y_5, y_5', y_6, y_6'\} \cup \{z_1, z_1', z_6, z_6'\}$$

For all $x \in X$ and $z \in Z'$, define

$$Q'(z \mid x) = \begin{cases} W(y_1 \mid x) + t(\alpha_1, \beta_1 \mid x), z = z_1 \\ W(y_1' \mid x) + t(\alpha_1, \beta_1 \mid x), z = z_1' \\ W(y_6 \mid x) + t(\alpha_6, \beta_6 \mid x), z = z_6 \\ W(y_6' \mid x) + t(\alpha_6, \beta_6 \mid x), z = z_6' \\ W(z \mid x), \text{ otherwise} \end{cases}$$

Then $Q' \geq W$, that is, $Q'$ is upgraded with respect to $W$.

∎

*Proof:* From Definition 2, for $Q'$ to be upgraded with respect to $W$, an intermediate channel $P: Z' \to Y$ must be found such that

$$W(Y \mid X) = Q'(Z' \mid X) P(Y \mid Z') \qquad (V)$$

The transition probability matrix for $W$ is

$$W(Y|X) = \begin{pmatrix} a_1 & a_2 & a_3 & a_4 & a_5 & a_6 & b_6 & b_5 & b_4 & b_3 & b_2 & b_1 \\ b_1 & b_2 & b_3 & b_4 & b_5 & b_6 & a_6 & a_5 & a_4 & a_3 & a_2 & a_1 \end{pmatrix}$$

Considering the definition of $Q'$ in the lemma, the transition probability matrix for $Q'$ can be expressed as

$$Q'(Z'|X) = \begin{pmatrix} a_1+\alpha_1 & a_6+\alpha_6 & b_6+\alpha_6 & b_1+\alpha_6 \\ b_1+\beta_1 & a_6+\beta_6 & b_6+\beta_6 & a_1+\beta_1 \end{pmatrix}$$

$Q'$ is a 2×4 matrix and $W$ is a 2×12 matrix. Thus, according to (V), the transition probability matrix for the intermediate channel $P$ must be a 4×12 matrix and in general this matrix can be assumed to be

$$P(Y|Z') = \begin{pmatrix} p_1 & p_2 & p_3 & p_4 & p_5 & p_6 & 0 & 0 & 0 & 0 & 0 & 0 \\ q_1 & q_2 & q_3 & q_4 & q_5 & q_6 & 0 & 0 & 0 & 0 & 0 & 0 \\ 0 & 0 & 0 & 0 & 0 & 0 & q_6 & q_5 & q_4 & q_3 & q_2 & q_1 \\ 0 & 0 & 0 & 0 & 0 & 0 & p_6 & p_5 & p_4 & p_3 & p_2 & p_1 \end{pmatrix}$$

Substituting this matrix into (V) gives the following equations that can be used to determine $p_1, p_2, p_3, p_4, p_5, p_6, q_1, q_2, q_3, q_4, q_5, q_6$

$$a_1 = (a_1+\alpha_1)p_1 + (a_6+\alpha_6)q_1$$
$$b_1 = (b_1+\beta_1)p_1 + (a_6+\beta_6)q_1$$
$$b_1 = (b_6+\alpha_6)q_1 + (b_1+\alpha_6)p_1$$
$$a_1 = (b_6+\beta_6)q_1 + (a_1+\beta_1)p_1$$
$$a_2 = (a_1+\alpha_1)p_2 + (a_6+\alpha_6)q_2$$
$$b_2 = (b_1+\beta_1)p_2 + (a_6+\beta_6)q_2$$
$$b_2 = (b_6+\alpha_6)q_2 + (b_1+\alpha_6)p_2$$
$$a_2 = (b_6+\beta_6)q_2 + (a_1+\beta_1)p_2$$
$$a_3 = (a_1+\alpha_1)p_3 + (a_6+\alpha_6)q_3$$
$$b_3 = (b_1+\beta_1)p_3 + (a_6+\beta_6)q_3$$
$$b_3 = (b_6+\alpha_6)q_3 + (b_1+\alpha_6)p_3$$
$$a_3 = (b_6+\beta_6)q_3 + (a_1+\beta_1)p_3$$

$$a_4 = (a_1 + \alpha_1)p_4 + (a_6 + \alpha_6)q_4$$
$$b_4 = (b_1 + \beta_1)p_4 + (a_6 + \beta_6)q_4$$
$$b_4 = (b_6 + \alpha_6)q_4 + (b_1 + \alpha_6)p_4$$
$$a_4 = (b_6 + \beta_6)q_4 + (a_1 + \beta_1)p_4$$
$$a_5 = (a_1 + \alpha_1)p_5 + (a_6 + \alpha_6)q_5$$
$$b_5 = (b_1 + \beta_1)p_5 + (a_6 + \beta_6)q_5$$
$$b_5 = (b_6 + \alpha_6)q_5 + (b_1 + \alpha_6)p_5$$
$$a_5 = (b_6 + \beta_6)q_5 + (a_1 + \beta_1)p_5$$
$$a_6 = (a_1 + \alpha_1)p_6 + (a_6 + \alpha_6)q_6$$
$$b_6 = (b_1 + \beta_1)p_6 + (a_6 + \beta_6)q_6$$
$$b_6 = (b_6 + \alpha_6)q_6 + (b_1 + \alpha_6)p_6$$
$$a_6 = (b_6 + \beta_6)q_6 + (a_1 + \beta_1)p_6$$

Solving these equations gives

$$p_1 = a_1 / (a_1 + \alpha_1) = b_1 / (b_1 + \beta_1) = b_1 / (b_6 + \alpha_6) = a_1 / (b_6 + \beta_6)$$
$$p_2 = a_2 / (a_1 + \alpha_1) = b_2 / (b_1 + \beta_1) = b_2 / (b_6 + \alpha_6) = a_2 / (b_6 + \beta_6)$$
$$q_1 = q_2 = 0$$
$$q_3 = a_3 / (a_6 + \alpha_6) = b_3 / (a_6 + \beta_6) = b_3 / (b_6 + \alpha_6) = a_3 / (b_6 + \beta_6)$$
$$q_4 = a_4 / (a_6 + \alpha_6) = b_4 / (a_6 + \beta_6) = b_4 / (b_6 + \alpha_6) = a_4 / (b_6 + \beta_6)$$
$$p_3 = p_4 = 0$$
$$p_5 = a_5 / (a_1 + \alpha_1) = b_5 / (b_1 + \beta_1) = b_5 / (b_1 + \alpha_6) = a_5 / (a_1 + \beta_1)$$
$$q_5 = 0$$
$$q_6 = a_6 / (a_6 + \alpha_6) = b_6 / (a_6 + \beta_6) = b_6 / (b_6 + \alpha_6) = a_6 / (b_6 + \beta_6)$$
$$p_6 = 0$$

The resulting transition probability matrix $P$ satisfies the conditions to be a channel, which completes the proof that $Q'$ is upgraded with respect to $W$.

**Lemma 6:** Let $W, y_1, y_2, y_3, y_4, y_5$ and $y_6$ be as in Lemma 5. Denote by $Q'_{123456} : X \to Z'_{123456}$ the result of applying Lemma 5 to $W, y_1, y_2, y_3, y_4, y_5$ and $y_6$. Next, denote by $Q'_{23456} : X \to Z'_{23456}$ the result of applying Lemma 3 to $W, y_2, y_3, y_4, y_5$ and $y_6$. Then $Q'_{23456} \geq Q'_{123456} \geq W$.

∎

***Proof:*** From Definition 1, for $Q'_{123456}$ to be degraded with respect to $Q'_{23456}$, an intermediate channel $P: Z'_{23456} \to Z'_{123456}$ must be found such that

$$Q'_{123456}(Z_{123456} \mid X) = Q'_{23456}(Z_{23456} \mid X) P(Z_{123456} \mid Z_{23456}) \qquad \text{(VI)}$$

Recall that the two alphabets $Z'_{123456}$ and $Z'_{23456}$ satisfy

$$Z'_{123456} = \{z_1, z_6, z_1{}', z_6{}'\} \cup A$$

$$Z'_{23456} = \{z_2, z_6, z_2{}', z_6{}', y_1, y_1{}'\} \cup A$$

where $A = Y \setminus \{y_1, y_1{}', y_2, y_2{}', y_3, y_3{}', y_4, y_4{}', y_5, y_5{}', y_6, y_6{}'\}$ is the set of symbols not participating in either merge operation.

As determined previously, the transition probability matrix for the channel $Q'_{123456}$ is

$$Q'_{123456}(Z_{123456} \mid X) = \begin{pmatrix} a_1 + \alpha_1 & a_6 + \alpha_6 & b_6 + \alpha_6 & b_1 + \alpha_6 \\ b_1 + \beta_1 & a_6 + \beta_6 & b_6 + \beta_6 & a_1 + \beta_1 \end{pmatrix}$$

From the definition of $Q'_{23456}$ given in Lemma 3, the transition probability matrix for channel $Q'_{23456}$ is

$$Q'_{23456}(Z_{23456} \mid X) = \begin{pmatrix} a_2 + \alpha'_2 & a_6 + \alpha'_6 & a_1 & b_1 & b_6 + \alpha'_6 & b_2 + \alpha'_6 \\ b_2 + \beta'_2 & a_6 + \beta'_6 & b_1 & a_1 & b_6 + \beta'_6 & a_2 + \beta'_2 \end{pmatrix}$$

$Q'_{123456}$ is a 2×4 matrix and $Q'_{23456}$ is a 2×6 matrix. Thus, according to (VI), the transition probability matrix for the intermediate channel $P$ must be a 6×4 matrix. In general this matrix can be assumed to be

$$P(Z_{123456} \mid Z_{23456}) = \begin{pmatrix} p & 0 & 0 & 0 \\ 0 & 1 & 0 & 0 \\ q & 0 & 0 & 0 \\ 0 & 0 & 0 & q \\ 0 & 0 & 1 & 0 \\ 0 & 0 & 0 & p \end{pmatrix}$$

Substituting this result into (VI) gives the following equations from which $p$ and $q$ can be determined

$$a_1 + \alpha_1 = (a_2 + \alpha'_2)p + qa_1$$
$$a_6 + \alpha_6 = a_6 + \alpha'_6$$
$$b_6 + \alpha_6 = b_6 + \alpha'_6$$
$$b_1 + \alpha_6 = (b_2 + \alpha'_6)p + qb_1$$
$$b_1 + \beta_1 = (b_2 + \beta'_2)p + qb_1$$
$$a_6 + \beta_6 = a_6 + \beta'_6$$
$$b_6 + \beta_6 = b_6 + \beta'_6$$
$$a_1 + \beta_1 = (a_2 + \beta'_2)p + qa_1$$

Note that $\alpha'_2, \beta'_2, \alpha'_6$ and $\beta'_6$ are as defined in Lemma 3, which are

$$\alpha'_2 = \lambda_2(\lambda_6 b - a)/(\lambda_6 - \lambda_2)$$
$$\beta'_2 = (\lambda_6 b - a)/(\lambda_6 - \lambda_2)$$
$$\alpha'_6 = \lambda_6(a - \lambda_2 b)/(\lambda_6 - \lambda_2)$$
$$\beta'_6 = (a - \lambda_2 b)/(\lambda_6 - \lambda_2)$$
$$\alpha_4 = \lambda_4(a_3 + b_3)/(\lambda_4 + 1)$$
$$\beta_4 = (a_3 + b_3)/(\lambda_4 + 1)$$
$$a = a_4 + a_5 + \alpha_4$$
$$b = b_4 + b_5 + \beta_4$$

Solving these equations gives

$$p = \frac{a_1 + \alpha_1}{a_2 + \alpha'_2} = \frac{b_1 + \alpha_6}{b_2 + \alpha'_6} = \frac{b_1 + \beta_1}{b_2 + \beta'_2} = \frac{a_1 + \beta_1}{a_2 + \beta'_2}$$
$$q = 0$$

As before, choosing $p = 1$ results in a transition probability matrix that satisfies the requirements to be a channel, which completes the proof that $Q'_{123456}$ is degraded with respect to $Q'_{23456}$ and so is a better approximation for channel $W$.

### 2.1 Generalization

An examination of Lemmas 1 to 6 reveals a generalization of the merging procedure. In Lemma 1, 4 symbols were merged to obtain a better upgraded version of the channel than by merging fewer symbols. The process followed was to first merge the 2 symbols in the middle (using Tool 1) to degrade the original channel, and then merge the 3 remaining symbols (using Tool 3) to obtain an upgraded version of the channel. Lemma 2 proved that this results in a better approximation of the original channel compared to the upgraded channels in [4].

In Lemma 3, 5 symbols were merged, and Lemma 4 proved that this result is a better upgraded version of the channel compared to that obtained via Lemma 1. The merging process for this case has 3 steps. First 2 symbols in the middle are merged (using Tool 2) to obtain an upgraded version of the original channel. In the next step, 2 symbols in the middle of the 4 remaining symbols from step 1 are merged to degrade the new channel (using Tool 1). Finally, the 3 remained symbols from step 2 are merged such that an upgraded version of the channel is obtained (using Tool 3).

Thus, to merge 4 symbols, we first **degrade** the channel and then **upgrade** it, whereas to merge 5 symbols, we first **upgrade** the channel, then **degrade** it and then **upgrade** it again. Therefore, as can be expected (and as proven in Lemma 5), in order to merge 6 symbols to get a better approximation of the channel, we have to first **degrade**, then **upgrade**, then **degrade** again, and then **upgrade** it a final time.

This shows that a generalization is indeed possible to merge $M$ symbols and get a better approximation of the original channel $W$. In the case $M$ an even integer, first degrade the channel, then upgrade it, and repeat these two operations until 2 symbols are obtained. In the case $M$ an odd integer, begin by upgrading the channel, then degrading it, and repeat these operations until 2 symbols remain. This is illustrated below.

**4 Symbols: (1) Degrade, (2) Upgrade**

**5 Symbols: (1) Upgrade, (2) Degrade, (3) Upgrade**

**6 Symbols: (1) Degrade, (2) Upgrade, (3) Degrade, (4) Upgrade**
.
.
.
**M=2k Symbols: (1) Degrade, (2) Upgrade, (3) Degrade, (4) Upgrade, (5) Degrade …**
**M=2k+1 Symbols: (1) Upgrade, (2) Degrade, (3) Upgrade, (4) Degrade, (5) Upgrade …**

Note that as $M$ tends to infinity, the upgraded channel resulting from this algorithm will be equivalent to the original channel $W$.

Finally, this section concludes with Lemmas 7 and 8 which express this generalization process in a more mathematical rigorous form.

***Lemma 7:*** Let $W: X \to Y$ be a BMS channel and let $y_1, y_2, y_3, ..., y_M$ be the symbols in the output alphabet $Y$. Denote $\lambda_1 = LR(y_1), \lambda_2 = LR(y_2), \lambda_3 = LR(y_3), ..., \lambda_M = LR(y_M)$. Assume that

$$1 \leq \lambda_1 < \lambda_2 \leq \lambda_3 \leq \ldots \leq \lambda_{M-1} < \lambda_M .$$

Next, let

$$a_2 = W(y_2 \mid 0), b_2 = W(y_2' \mid 0), a_3 = W(y_3 \mid 0), b_3 = W(y_3' \mid 0),$$
$$\vdots$$
$$a_{M-2} = W(y_{M-2} \mid 0), b_{M-2} = W(y_{M-2}' \mid 0), a_{M-1} = W(y_{M-1} \mid 0), b_{M-1} = W(y_{M-1}' \mid 0)$$

In the case $M$ an even integer, take $y_2$ and $y_3$ and merge them together using Tool 1 to degrade the channel. In the next step, merge $y_{23}$ and $y_4$ using Tool 2 to upgrade the channel. Continue degrading and upgrading the channel respectively by merging the new symbol resulting from the previous step with the next consecutive symbol until 3 symbols remain $y_1, y_{2,3,\ldots,M-1}, y_M$. Then merge these 3 symbols using Tool 3 to upgrade the channel a final time.

In the case $M$ an odd integer, take $y_2$ and $y_3$ and merge them together using Tool 2 to upgrade the channel. In the next step, merge $y_{23}$ and $y_4$ using Tool 1 to degrade the channel. Continue upgrading and degrading the channel respectively by merging the new symbol resulting from the previous step with the next consecutive symbol until 3 symbols remain $y_1, y_{2,3,\ldots,M-1}, y_M$. Then merge these 3 symbols using Tool 3 to upgrade the channel a final time.

For real numbers $\alpha, \beta$ and $x \in X$, define

$$t(\alpha, \beta \mid x) = \begin{cases} \alpha, x = 0 \\ \beta, x = 1 \end{cases}$$

Define the BMS channel $Q': X \to Z'$ as follows.

The output alphabet $Z'$ is given by

$$Z' = Y \setminus \{y_1, y_1', y_2, y_2', \ldots, y_M, y_M'\} \cup \{z_1, z_1', z_M, z_M'\}$$

For all $x \in X$ and $z \in Z'$, define

$$Q'(z \mid x) = \begin{cases} W(y_1 \mid x) + t(\alpha_1, \beta_1 \mid x), z = z_1 \\ W(y_1' \mid x) + t(\alpha_1, \beta_1 \mid x), z = z_1' \\ W(y_M \mid x) + t(\alpha_M, \beta_M \mid x), z = z_M \\ W(y_M' \mid x) + t(\alpha_M, \beta_M \mid x), z = z_M' \\ W(z \mid x), \text{ otherwise} \end{cases}$$

Then $Q' \geq W$, that is, $Q'$ is upgraded with respect to $W$.

∎

**Proof:** From Definition 2, for $Q'$ to be upgraded with respect to $W$, an intermediate channel $P: Z' \to Y$ must be found such that
$$W(Y|X) = Q'(Z'|X)P(Y|Z') \qquad \text{(VII)}$$

The transition probability matrix for $W$ is

$$W(Y|X) = \begin{pmatrix} a_1 & a_2 & a_3 & \cdots & a_{M-1} & a_M & b_M & b_{M-1} & \cdots & b_3 & b_2 & b_1 \\ b_1 & b_2 & b_3 & \cdots & b_{M-1} & b_M & a_M & a_{M-1} & \cdots & a_3 & a_2 & a_1 \end{pmatrix}$$

Considering the definition of $Q'$ in the lemma, the transition probability matrix for $Q'$ can be expressed as

$$Q'(Z'|X) = \begin{pmatrix} a_1 + \alpha_1 & a_M + \alpha_M & b_M + \alpha_M & b_1 + \alpha_M \\ b_1 + \beta_1 & a_M + \beta_M & b_M + \beta_M & a_1 + \beta_1 \end{pmatrix}$$

$Q'$ is a $2 \times 4$ matrix and $W$ is a $2 \times 2M$ matrix. Thus, according to (VII), the transition probability matrix for the intermediate channel $P$ must be a $4 \times 2M$ matrix, and in general this matrix can be assumed to be

$$P(Y|Z') = \begin{pmatrix} p_1 & p_2 & \cdots & p_M & 0 & \cdots & 0 & 0 \\ q_1 & q_2 & \cdots & q_M & 0 & \cdots & 0 & 0 \\ 0 & 0 & \cdots & 0 & q_M & \cdots & q_2 & q_1 \\ 0 & 0 & \cdots & 0 & p_M & \cdots & p_2 & p_1 \end{pmatrix}$$

Substituting this matrix into (VII) gives the following equations that can be used to determine $p_1, p_2, \ldots, p_M, q_1, q_2, \ldots, q_M$

$$a_1 = (a_1 + \alpha_1)p_1 + (a_M + \alpha_M)q_1$$
$$b_1 = (b_1 + \beta_1)p_1 + (a_M + \beta_M)q_1$$
$$b_1 = (b_M + \alpha_M)q_1 + (b_1 + \alpha_M)p_1$$
$$a_1 = (b_M + \beta_M)q_1 + (a_1 + \beta_1)p_1$$
...
...
...
$$a_M = (a_1 + \alpha_1)p_M + (a_M + \alpha_M)q_M$$
$$b_M = (b_1 + \beta_1)p_M + (a_M + \beta_M)q_M$$
$$b_M = (b_M + \alpha_M)q_M + (b_1 + \alpha_M)p_M$$
$$a_M = (b_M + \beta_M)q_M + (a_1 + \beta_1)p_M$$

Solving these equations gives

$$p_1 = a_1/(a_1 + \alpha_1) = b_1/(a_1 + \beta_1) = b_1/(b_1 + \alpha_M) = a_1/(a_1 + \beta_1)$$
$$q_1 = 0$$
$$p_2 = 0$$
$$q_2 = a_2/(a_M + \alpha_M) = b_2/(a_M + \beta_M) = b_2/(b_M + \alpha_M) = a_2/(b_M + \beta_M)$$
...
...
...
$$p_{(M=2k)} = 0$$
$$q_{(M=2k)} = a_M/(a_M + \alpha_M) = b_M/(a_M + \beta_M) = b_M/(b_M + \alpha_M) = a_M/(b_M + \beta_M)$$
...
$$q_{(M=2k+1)} = 0$$
$$p_{(M=2k+1)} = a_M/(a_1 + \alpha_1) = b_M/(a_1 + \beta_1) = b_M/(b_1 + \alpha_M) = a_M/(a_1 + \beta_1)$$

The resulting transition probability matrix $P$ satisfies the conditions to be a channel, which completes the proof that $Q'$ is upgraded with respect to $W$.

***Lemma 8:*** Let $W, y_1, y_2, y_3, ..., y_M$ be as in Lemma 7. Denote by $Q'_{123...M} : X \to Z'_{123...M}$ the result of applying Lemma 7 to $W, y_1, y_2, y_3, ..., y_M$. Next, denote by $Q'_{23...M} : X \to Z'_{23...M}$ the result of applying Lemma 7 for $N = (M-1)$ symbols to $W, y_2, y_3, ..., y_M$. Then $Q'_{23...M} \geq Q'_{123...M} \geq W$.

∎

**Proof:** From Definition 1, for $Q'_{123...M}$ to be degraded with respect to $Q'_{23...M}$, an intermediate channel $P: Z'_{234...M} \to Z'_{1234...M}$ must be found such that

$$Q'_{123...M}(Z_{1234...M} | X) = Q'_{23...M}(Z_{234...M} | X) P(Z_{1234...M} | Z_{234...M}) \quad \text{(VIII)}$$

Recall that the two alphabets $Z'_{1234...M}$ and $Z'_{234...M}$ satisfy

$$Z'_{1234...M} = \{z_1, z_M, z_1', z_M'\} \cup A$$

$$Z'_{234...M} = \{z_2, z_m, z_2', z_M', y_1, y_1'\} \cup A$$

where $A = Y \setminus \{y_1, y_1', y_2, y_2', ..., y_M, y_M'\}$ is the set of symbols not participating in either merge operation.

As determined previously, the transition probability matrix for the channel $Q'_{123...M}$ is

$$Q'_{123...M}(Z_{1234...M} | X) = \begin{pmatrix} a_1 + \alpha_1 & a_M + \alpha_M & b_M + \alpha_M & b_1 + \alpha_M \\ b_1 + \beta_1 & a_M + \beta_M & b_M + \beta_M & \alpha_1 + \beta_1 \end{pmatrix}$$

From the definition of $Q'_{23...M}$ given in Lemma 7, the transition probability matrix for channel $Q'_{23...M}$ is

$$Q'_{23...M}(Z_{234...M} | X) = \begin{pmatrix} a_2 + \alpha'_2 & a_M + \alpha'_M & a_1 & b_1 & b_M + \alpha'_M & b_2 + \alpha'_M \\ b_2 + \beta'_2 & a_M + \beta'_M & b_1 & a_1 & b_M + \beta'_M & a_2 + \beta'_2 \end{pmatrix}$$

$Q'_{123...M}$ is a 2×4 matrix and $Q'_{23...M}$ is a 2×6 matrix. Thus, according to (VIII), the transition probability matrix for the intermediate channel $P$ must be a 6×4 matrix. In general this matrix can be assumed to be

$$P(Z_{1234...M} | Z_{234...M}) = \begin{pmatrix} p & 0 & 0 & 0 \\ 0 & 1 & 0 & 0 \\ q & 0 & 0 & 0 \\ 0 & 0 & 0 & q \\ 0 & 0 & 1 & 0 \\ 0 & 0 & 0 & p \end{pmatrix}$$

Substituting this result into (VIII) gives the following equations from which $p$ and $q$ can be determined

$$a_1 + \alpha_1 = (a_2 + \alpha'_2)p + qa_1$$
$$a_M + \alpha_M = a_M + \alpha'_M$$
$$b_M + \alpha_M = b_M + \alpha'_M$$
$$b_1 + \alpha_M = (b_2 + \alpha'_M)p + qb_1$$
$$b_1 + \beta_1 = (b_2 + \beta'_2)p + qb_1$$
$$a_M + \beta_M = a_M + \beta'_M$$
$$b_M + \beta_M = b_M + \beta'_M$$
$$a_1 + \beta_1 = (a_2 + \beta'_2)p + qa_1$$

Note that $\alpha'_2, \beta'_2, \alpha'_M$ and $\beta'_M$ are as defined in Lemma 7.

Solving these equations gives

$$p = \frac{a_1 + \alpha_1}{a_2 + \alpha'_2} = \frac{b_1 + \alpha_M}{b_2 + \alpha'_M} = \frac{b_1 + \beta_1}{b_2 + \beta'_2} = \frac{a_1 + \beta_1}{a_2 + \beta'_2}$$
$$q = 0$$

As before, choosing $p = 1$ results in a transition probability matrix that satisfies the requirements to be a channel, which completes the proof that $Q'_{123...M}$ is degraded with respect to $Q'_{23...M}$ and so is a better approximation for channel $W$.

## 3. Conclusion

In the section, a summary of the results presented in this paper is presented, followed by some complementary remarks.

Section 1 provided an introduction to polar codes and the approaches that have been taken to provide a straightforward means of constructing them. The recent and successful approach by Tal and Vardy was then described, and their concepts explained in detail for use later in the paper. At each stage of polarization, a power of two bit-channels are obtained. As the channel polarization process continues and the number of bit-channels increases. The growth in the output alphabet size of the bit-channels is extreme, so that using the definitions for code construction is computationally very complex. Thus an approximation is used. A desired output alphabet size is assumed and then all bit-channel output alphabets are reduced to this value while being transformed into two different channels. The first is degraded with respect to the original channel, i.e., has a lower rate, and the second is upgraded with respect to the original channel, i.e., has a higher rate. These two channels are achieved by merging output symbols. Degrading and upgrading merging functions are introduced as tools for code construction. Then a method for constructing polar codes is presented based on the degraded channel. The distance from the optimal case is determined by constructing codes using the upgraded channels.

In Section 2, the tools for code construction are used to expand on the concepts in Section 1. The need to consider the degraded channel and the corresponding code construction is eliminated. First, an algorithm is introduced that merges 4 consecutive output symbols so that the resulting upgraded channel is closer to the original channel than previous techniques. This approach is then extended to merging 5 and 6 consecutive output symbols. Merging larger numbers of symbols is achieved by using iterative upgrading and degrading. The resulting channel is arbitrarily close to the original bit-channel and therefore to the optimal construction of a polar code from the perspective of the upgraded channel. The generalization to an arbitrary number of consecutive output symbols being merged together results in a channel equivalent to the original channel and therefore, optimal construction of polar codes is achieved.